\documentclass[]{aa}
\usepackage{graphicx}
\begin{document}
\title{V39: an unusual object in the field of IC 1613
\thanks{Based in part on observations collected at ESO--La Silla}}
\author{L. Mantegazza\inst{1} \and E. Antonello\inst{1} \and  
D. Fugazza\inst{1}\and S. Covino\inst{1}  \and G. Israel\inst{2}
 }
\institute{Osservatorio Astronomico di Brera, Via E.~Bianchi 46,
       I--23807 Merate, Italy \and
Osservatorio Astronomico di Roma, Via di Frascati 33, I-00040
       Monteporzio Catone, Italy}
\offprints{L. Mantegazza \\ \email{luciano@merate.mi.astro.it}}
\date{ Received date; accepted date }
\titlerunning{V39: an unusual star}
\authorrunning{L. Mantegazza et al.}

\abstract
{The variable star V39 in the field of IC 1613 is discussed in the light
of 
the available photometric and new spectroscopic data. It has 
strong emission Balmer lines, and the observed characteristics could be 
explained by a W Vir pulsating star with a period of 14.341 d, located
at 
more than 115 kpc, that is in the very outer halo of our Galaxy. 
It should have an apparent companion, a long period (1118d) red variable, 
belonging to IC 1613. The main uncertainty in this interpretation is
an emission feature at 6684 \AA, which we tentatively identified as a 
He I line. 
\keywords{Stars: emission-line -- Stars: variables: general -- Galaxy: halo
 -- Galaxy: structure}
}
\maketitle

\section{Introduction}
The variable star V39, $\alpha(2000)$=$1^h 05^m 02^s.1$, 
$\delta(2000)$=$2^o 10'24''$, 
was discovered by Sandage (1971) in the field of IC 1613. The
photographic 
$B$ observations, with an average apparent magnitude of $B=19.2$,  
showed a light curve with the shape of an inverted $\beta$ Lyr eclipsing 
variable with a period of 28.72 days. Sandage 
included it among the possible Cepheids, but in the subsequent 
works (see e.g. Madore \& Freedman, 1981) it was never used for deriving 
the $PL$ relation of 
Cepheids. Hutchinson (1973) re-examined these observations and suggested
it 
could be a W Vir star located very far in the halo of our Galaxy.
Van den Bergh (2000) suggested that it could be an isolated star in 
the intergalactic space. Hutchinson noted also a long term variability
of 
about 1000 d. Antonello et al. (1999) surveyed IC 1613 looking for
Cepheids, 
and added further data points to the time series of V39; the data, taken
in 
unfiltered light, seemed to confirm that this is not a classical
pulsating 
star, however its real nature remained unclear. The new observations
indicated a long $P$ of 1123 d and as regards the shorter one it was not 
possible to select unambiguously between 28.699 d and half this value, 
14.350 d. New $V$ and $I$ observations were obtained in the context of
the 
OGLE project (Udalski et al. 2001), but these data alone, due to their 
short baseline, are insufficient to clarify the matter. In 1999 some
spectra 
were taken at ESO-LaSilla, in order to throw new light on the nature of
this 
object; in the present note we rediscuss all the photometric data 
and report on the analysis of the spectra.

\section{Photometry}

The available photometry includes: 1) the $B$ photographic data
published 
by Sandage (1971, 103 datapoints), which were derived from plates taken
by 
Baade from 1929 to 1937; 2) our unfiltered observations ($Wh$ band,
Antonello 
et al. 1999, 66 datapoints) taken from 1995 to 1998, with further 6  
datapoints obtained by E. Poretti with the 1.5 m telescope of San Pedro Martir 
Observatory in 1999; 3) the OGLE $VI$ observations obtained by Udalski
et 
al. (2001) in 2000 (40 and 42 datapoints, respectively). Another two 
datapoints were recovered: one $BVRI$ observation obtained by Freedman 
(1988) and one $VR$ measurement obtained by us with the Dutch telescope 
(Antonello et al. 1999). These last data are listed in Table 1 along
with
the mean values of OGLE data.
\begin{table}
\centering
\caption[]{Colors of V39}
\begin{tabular}{ccccc}
\hline
J.D. &V &B-V& V-R& V-I\\
 (2400000+)& & & & \\
\hline
45973.02&  18.59 & 0.56 & 0.59 & 1.12 \\
50302.87&  18.48 & ---  & 0.51 & ---  \\
51830.  &  18.85 & ---  & ---  & 1.23 \\
\hline
\end{tabular}
\end{table}
\begin{figure}
 \resizebox{\hsize}{!}{\includegraphics{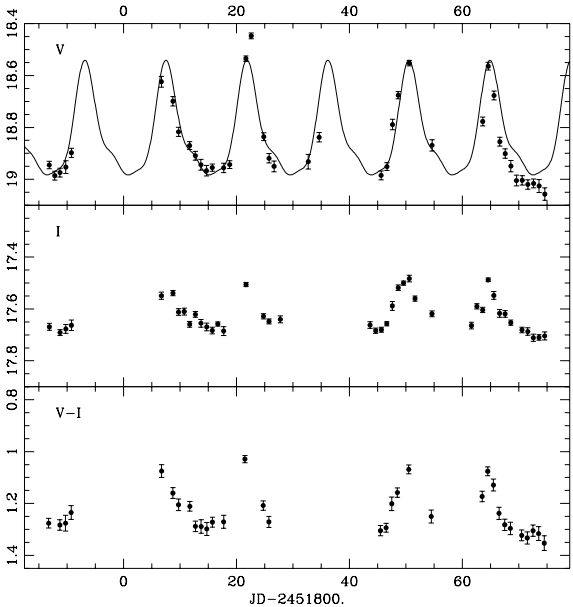}}
 \caption[ ]{OGLE $V$,$I$,$V-I$ observations. The solid curve in the top 
panel is the best fitting curve with P=14.3411d.} 
\end{figure}
Fig. 1 shows the OGLE $V,I,V-I$ data plotted vs. the time. We can see
that 
they present a clear variation with a period of about 14 days and a sort
of irregularity in the light curve. 
The three panels have the same scale so that it is possible to 
appreciate the different amplitudes of the curves ($\Delta V\sim0.45$, 
$\Delta I\sim0.2$ mag).

In order to get profit of the whole available information we decided to
merge all the datasets by transforming them to a common system.
Therefore both the $B$-photographic and $Wh$ data were shifted and
rescaled to match the $V$ data (for the transformation of $Wh$ colour to the 
$V$ one see Antonello et al. 1999 and also Riess et al. 1999; Sandage's
$B$ data were transformed assuming that there have not been amplitude 
variations). 
When doing this we neglected the small phase shifts between the different 
color curves, because they are uninmportant for the period search. 
As a result we obtained a dataset of 219 mesurements spanning about 26000 d.
If we plot these data vs time we see that there is a long--period
variation.
Its presence has been already suggested by Hutchinson (1970)
on the basis of Sandage's data alone, an confirmed by Antonello et al. 
(1999) on the basis of their $Wh$ data.

We frequency analized the data with the least
squares 
power spectrum technique developed to study multiperiodic signal with 
unequally spaced data (Vanicek, 1971; Antonello et al., 1986). We detected 
3 periodic 
components, a long period term with $P_L=1118$ d ($\nu_L=8.9\,10^{-4}$ 
d$^{-1}$)
, a short period one with 
$P_S=14.3411$d ($\nu_S=0.06973 d^{-1}$) and its first harmonic. 
The first two power spectra which show the two dominat periods are shown
in the top panel of Fig.2. The bottom panel of the same figure shows the
spectral window, which, beside the presence of the peak at zero frequency with
its one reciprocal year sidelobes, shows another peak at $\nu_w=0.0339 d^{-1}$
(reciprocal of the synodic month), flanked by its  one reciprocal year 
sidelobes.
The double-wave period of 28.677 d is an artifact generated by this structure 
of the spectral window, in fact its frequency ( 0.0348d$^{-1}$) corresponds
to $\nu_L+\nu_w$ (i.e. it is an alias of the long period). It is a mere 
coincidence that this value is almost exactly twice the short period.
The $V$ amplitude of the short period variation is about 
0.45 mag, while that of the long--period one is about 0.40 mag.
 After removing the long and short period
variations, it is apparent that the data dispersion is still large,
0.096 mag, to be compared with the estimated mean white noise level of 0.068
mag. 
If we push further on the analysis we find another peak at 
$\nu_4=0.07050 d^{-1}$, which satisfies  the relation 
$\nu_4\simeq \nu_S+\nu_L$. This relation could indicate that this 
is a non-linear coupling term between $\nu_L$ and $\nu_S$, suggesting 
a physical connection between the two periodic variations. However 
this is a spurious term introduced by the fact that we consider the 
data in magnitudes; the non--linear transformation from intensities to 
magnitudes introduces an apparent modulation of the amplitude of the
short--period term. In fact, when analyzing the intensity data, the term at 
0.0705$d^{-1}$ is not present.
\begin{figure}
 \resizebox{\hsize}{!}{\includegraphics{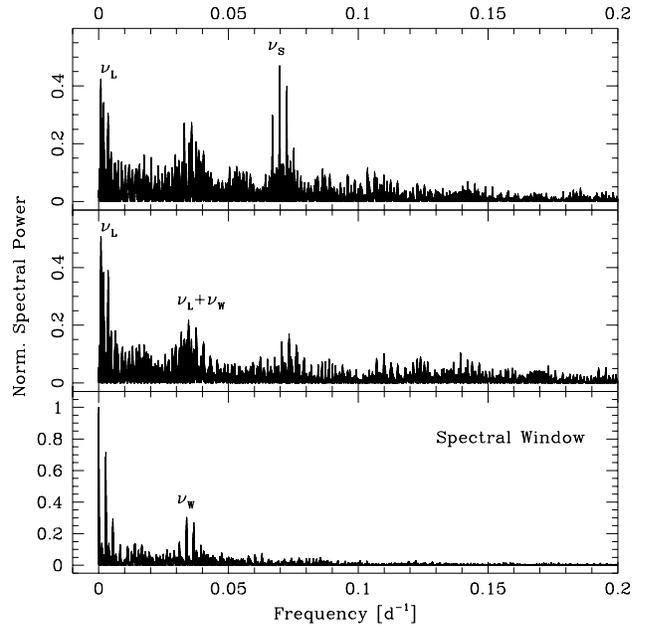}}
 \caption[ ]{Upper panel: original least-squares power spectrum; central panel:
power spectrum after introducing $\nu_S$ as a known constituent, bottom
panel: spectral window. The labelled peaks are discussed in the text} 
\end{figure}
Figure 3 shows all the available data after removing the long 
period variation and the amplitude modulation due to it, phased with 
the period of 14.341 d. The resulting light curve has the typical shape 
of that of a pulsating variable, even if the dispersion of the data is
still larger than that expected from data errors; probably the light curve 
does not repeat exactly from cycle to cycle. This can be seen also in the
top
panel of Fig. 1, which shows the OGLE V data; for comparison the solid
line 
represents the best fit with the period of 14.3411 d.
\begin{figure}
 \resizebox{\hsize}{!}{\includegraphics{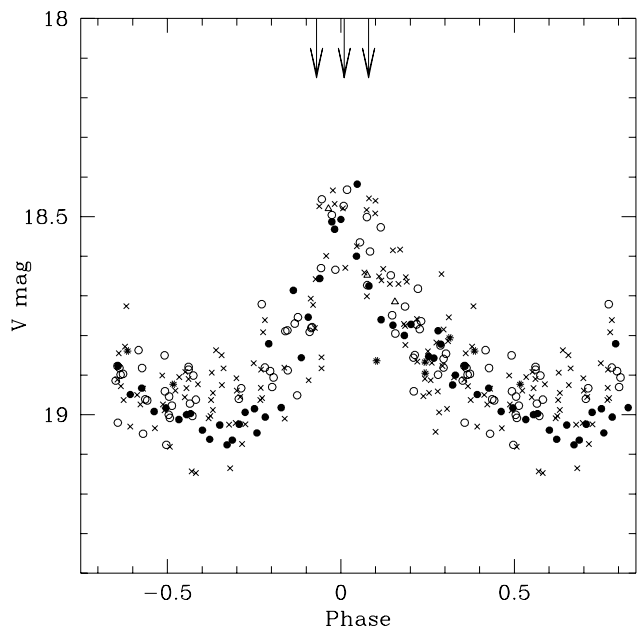}}
 \caption[ ]{Photometric data of V39 transformed into V
magnitudes 
and phased with the 14.341 d period after removing the long period 
variation. Crosses: Sandage (1970); open circles: Antonello et al.
(1999); 
asterisks: data taken at San Pedro Martir; closed circles: OGLE data. 
Arrows: phases of spectroscopic observations.}
\end{figure}
Fig. 4 shows the long period variation curve with $P=1118$ d, obtained
after 
removing the short-period variation. 
\begin{figure}
 \resizebox{\hsize}{!}{\includegraphics{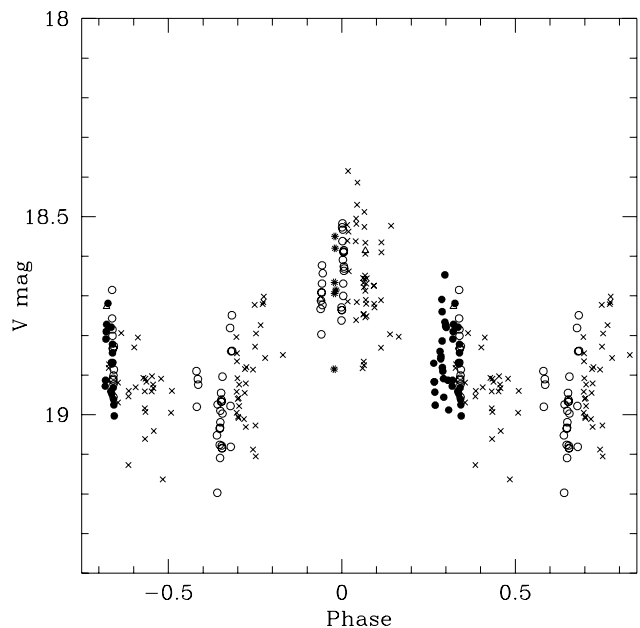}}
 \caption[ ]{Photometric data of V39 transformed into V
magnitudes 
and phased with the 1118 d period after removing the short--period
variation. 
Symbols as in Fig. 3}
\end{figure}

\section{Spectroscopy}

\subsection{Observations}

The spectroscopic observations were performed with the the Danish 1.54 m 
telescope
equipped with the DFOSC, and the EFOSC-2  attached to the
3.6 m telescope of the La Silla Astronomical Observatory (ESO) from 12 to 14 
September 1999. The observations with the DFOSC were taken with
the 
grism \#4, with a nominal resolution of $\Delta \lambda$ = 8.3 \AA~ and a
range
from 3500 to 7000 \AA; whereas the observations with the EFOSC-2 camera
were 
performed with the grism \#11 with a nominal resolution of 
$\Delta\lambda$=13.2 \AA~ and a range from 3380 to 7520 \AA. Bias and
twilight 
flat field frames were gathered in each night. A Helium lamp image was
taken 
soon afterwards the observations. During these three nights we collected
a 
total of 6 images, each with an exposure time of 1200 sec. The complete
log 
of observations is reported in Table 2. The table contains for each
image the 
date of the beginning of the night, the heliocentric Julian date of
midexposure, 
the airmass, the slit width in arcsec and the used camera.
\begin{table}
\centering
\caption []{Characteristics of the spectrograms}
\begin{tabular}{cccccc}
\hline
date & J.D. &Airmass&Slit& Instrument\\
&&&width\\
 &(2400000+)& & (") &\\
\hline
Sep. 12 & 51433.75 & 1.172 & 2.5 & DFOSC\\
Sep. 13 & 51434.84 & 1.389 & 2.0 & DFOSC\\
 & 51434.86 & 1.513 & 2.0 & DFOSC\\
 & 51434.88 & 1.686 & 2.0 & DFOSC\\
Sep. 14 & 51435.82 & 1.377 & 1.0 & EFOSC-2\\
 & 51435.84 & 1.308 & 1.0 & EFOSC-2\\
\hline
\end{tabular}
\end{table}
The phases of the spectra computed according to the ephemeris:
$T_{Max}(JD)=2451850.6+14.3411E$ are: -0.07 (Sep. 12, 1999), 0.01
(Sep. 13), 0.08 (Sep 14), i.e the observations were performed close to
the maximum brightness.

\subsection{Data reduction and Analysis}
The standard corrections and the cosmic-ray removal were performed using
the ESO/MIDAS packages. All spectra were wavelength calibrated and
extracted 
by means of the ESO/MIDAS (99NOV) ``long'' context.  

In order to improve the $S/N$ ratio the three spectra of Sept. 13 and the
two of Sept. 14 were coadded respectively. At the continuum level 
in the region around 6100\AA~ the mean Danish telescope 
spectra have a $S/N\sim30$ and those of the 3.6m telescope $S/N\sim50$.
The resolutions, as  measured from the width of the sky lines, are about
16\AA~ for the Danish telescope and 13\AA~ for the 3.6m telescope.
The single Danish spectrum of Sep. 12 has $S/N\sim23$, however it
is slightly defocussed (FWHM=23\AA).
The spectra of Sept. 13 and 14 are shown in Fig. 5, where the sky 
spectrum has been subtracted.
We can see
 four stellar emission lines,$ H\alpha$,$ H\beta$,$ H\gamma$, and a broad
feature at 6684 \AA~ (FWHM$\sim 24$\AA, to be compared to that of 
the adjacent 
$H\alpha$ of about 13\AA). We tentatively suggest that it could be the 
$HeI$ 6678 line; no other strong lines are usually present in the vicinity
of
this wavelength in stellar spectra. However, in this case the feature is 
red--shifted by about 180 kms$^{-1}$, while $H$ lines on the contrary
seem 
rather blue-shifted.  The feature is real because it is
present in both the 3.6m telescope spectrograms, while in the Danish 
telescope spectrograms it is barely detectable.
\begin{figure}
 \resizebox{\hsize}{!}{\includegraphics{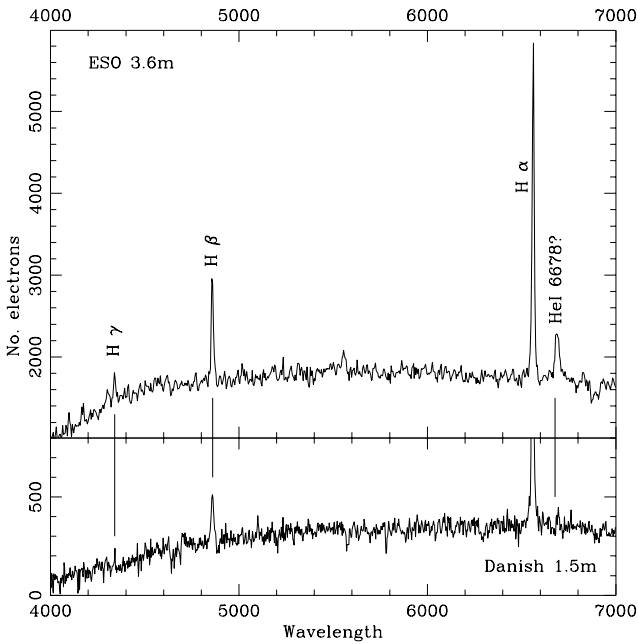}}
 \caption[ ]{Spectra of V39 after removal of the sky contribution. 
The stellar lines are labelled.} 
\end{figure}
Other weak emissions (in particular in the 3.6m telescope spectra)
seem to be present, but they are not clearly associable to well--known
spectral features (apart perhaps the $H\delta$) and, at least in
part, are spurious features due to the sky subtraction.
At the same time, no absorption features are 
unambiguously detectable.

We tried to verify the membership of V39 to the galaxy IC1613 by means 
of the radial velocities. No conclusive results were found because the 
$H$ lines supply discordant values.

\section{Discussion}
We can summarize the main observed characteristics of V39 as follows:
\begin{itemize}
\item There are light variations with a period of 14.341 d. The
corresponding
light curve has the typical shape of a Cepheid-like variable, even if it
is 
not perfectly regular. Its amplitude decrease with increasing wavelength
($\Delta{B}=0.9$, $\Delta{V}=0.45$, $\Delta{I}=0.2$, assuming no
significant 
amplitude change over seventy years). $V$ and $I$ variations are in
phase and 
the star is bluer at maximum luminosity.

\item There is a long--period variation with P=1118d and $V$ 
amplitude of about 0.35 mag.

\item The spectra at phases close to the maximum brigthness show strong 
$H$ lines emissions. At these phases a feature at $6684\AA~$ ($HeI$ 6678?)
is also visible in emission.

\end{itemize}

\subsection{V39 as a pulsating star}
While it seems very probable that the star is pulsating with a period of 
about 14 d, it is an unlikely classical pop I Cepheid. The presence of 
irregularities in the light curve and of $H$ and $HeI$(?) emission at
maximum 
light indicate that it could be a pop. II Cepheid (i.e.a W Vir
star;
Lebre \& Gillet, 1992). With this interpretation some questions still
remain 
open: a) one should expect also the emission line $HeI$ 5876\AA; however
due to  the modest resolution of our spectra, this line, if present, 
should be
blended, especially if red--shifted as the 6678\AA~ one, with the much
stronger 
sky emission line of $NaI$ 5890; b) although the presence of $H$ emission
lines 
at the observed phases is in agreement with the case of W Vir 
(Lebre \& Gillet, 1992), in such a case the $HeI$ emission should be most 
clearly seen between phases 0.61--0.003, while our best detection is at 
phase 0.08. May be this can be ascribed to the non-perfect regularity of
the 
variations which can affect the accurate estimate of the maximum phase 
(see Fig. 1), and to the fact that spectroscopic and
photometric 
observations are not simultaneous. If the star were a W Vir variable,
its apparent brightness would exclude its membership to IC1613, 
and it should be a foreground object.

The long period variations could be ascribed to the presence of a
variable red star, a real companion or more plausibly a background 
star belonging to IC1613 (
a not unlikely case, see the analogous example of the pop. I Cepheid 
$V2942B$, Antonello et al 2000). This is supported by the observed
color. 
The mean $V-I$ is 1.23 while typical values of W Vir (or metal poor
population
I Cepheids) are around 0.6. We cannot ascribe this discrepancy to the 
foreground interstellar reddening because it is negligible in the
direction 
of IC1613 ($<E(V-I)>=0.09$, Macri et al. 2001). A long--period red
variable 
with an apparent $V$ magnitude which is 2 mag fainter than the W Vir
star and 
with $V-I=2.5$ could approximately explain the observed colour index. 
Moreover it reduces the ratio between $V$ and $I$ amplitudes of the 14 d 
period variation from the observed value of about 2.3 to 1.4, which is 
similar to that of known W Vir stars. If this were the case, the long--period 
variation would have a real $V$ amplitude of about 1.4 mag, while the
$V$ 
amplitude of the short-period variation would increase only marginally. 
The red star should be a supergiant with $V\sim20.5$ and should belong 
to IC1613. Red variable stars with similar period and brigthness were
found 
by us in this galaxy (see for instance Table 9 of Antonello et al.,
2000).

\subsection{V39 as a high energy source?}
The strong emission lines and the photometric variations led us to
suspect that the star could be an unusual object belonging to the
class of cataclysmic 
variables, that is a relatively nearby star in a binary system with an
accreting disk or envelope around a compact object. However the result of the 
photometric analysis (in particular the ratio of the amplitudes in the 
different bands and the phase relations between $V$ and $I$)
rules out a geometric origin of the variability. 
Moreover, other spectral lines should have been observed. The intriguing 
feature is the line at 6684 \AA, which is difficult to explain in the 
context of the stellar pulsation, and reminded us of the features
observed
in an unusual object such as SS433. However, no X-ray source has been
found
in the catalogues at the location of V39. The comparison of the published
coordinates of the star suggests the lack of a proper
motion during the last seventy years, and this support the view that it
is a far object.

\section{Conclusion}
The observational data tend to indicate that V39 is a foreground W Vir
star
of IC1613. Adopting the $PL$ relation by Nemec et al. (1994), assuming a 
[Fe/H]=--1 and a fundamental mode pulsation we get $M_V=-1.7$, which for 
$<V_0>=18.63$ gives a lower limit of distance modulus of 20.3, while
the  
distance modulus of IC1613 is 24.5 (Macri et al., 2001). We also
speculated 
that the star could have a red companion, but its presence is not
sufficient 
to confirm the membership of V39 to IC1613. In fact in this case the
companion 
should be much brigther than the W Vir star, the amplitude of the 14 d 
period variation would be much smaller than what observed,
and the red companion contribution to the spectrum should be dominating.
In other words, while the W Vir star belongs to the halo of our Galaxy, 
the red star
should belong to IC 1613. The presence of an isolated W Vir star at a
distance
of at least 115 Kpc (assuming [Fe/H]=--2, the distance would be about 130 kpc)
should throw new fuel on the open question of the
true 
extent of the galactic halo. If true this would be the farthest
known star of our galaxy; known halo field stars and globular clusters are
closer than about 100 kpc (e.g. Morrison et al. 2001).
Therefore it should be very important to 
definitively settle the question on the nature of V39. In order to do
this, 
some spectra, at higher resolution of the present ones, should be taken
for
confirming the reality of the presence of $HeI$ emissions, and for
deriving 
accurate radial velocities, which could definitively exclude the
membership 
of IC1613. The spectra should be taken both at phases in which the 
emission lines are present in W Vir stars (the ascending branch of the 
ligth curve) and at phases of minimum light, where the emission lines
should be absent. 
\begin{acknowledgements}
We are grateful to Dr. E. Poretti for supplying us with the SPM $Wh$ band 
observations and for useful comments and suggestions.
\end{acknowledgements}

\end{document}